\documentclass[prl,superscriptaddress,twocolumn,amsmath]{revtex4-1}

\usepackage{bm}
\usepackage{graphicx}
\usepackage[usenames,dvipsnames,svgnames,table]{xcolor}
\usepackage[colorlinks=true,linkcolor=RoyalBlue,citecolor=RoyalBlue]{hyperref}

\newcommand{\eps}{\varepsilon}
\newcommand{\dpar}[2]{\frac{\partial #1}{\partial #2}}
\newcommand{\ket}[1]{\lvert #1 \rangle}
\newcommand{\bracket}[1]{\langle #1 \rangle}
\DeclareMathOperator{\IM}{Im}

\begin{document}

\title{Berry phase modification to the energy spectrum of excitons}

\author{Jianhui Zhou}
\affiliation{Department of Physics, Carnegie Mellon University, Pittsburgh, Pennsylvania 15213, USA
}

\author{Wen-Yu Shan}
\affiliation{Department of Physics, Carnegie Mellon University, Pittsburgh, Pennsylvania 15213, USA
}

\author{Wang Yao}
\affiliation{Department of Physics and Center of Theoretical and Computational Physics, University of Hong Kong, Hong Kong, China}

\author{Di Xiao}
\affiliation{Department of Physics, Carnegie Mellon University, Pittsburgh, Pennsylvania 15213, USA
}
%\date{\today}

\begin{abstract}
By quantizing the semiclassical motion of excitons, we show that the Berry curvature can cause an energy splitting between exciton states with opposite angular momentum.  This splitting is determined by the Berry curvature flux through the $\bm k$-space area spanned by the relative motion of the electron-hole pair in the exciton wave function.  Using the gapped two-dimensional Dirac equation as a model, we show that this splitting can be understood as an effective spin-orbit coupling effect.  In addition, there is also an energy shift caused by other ``relativistic'' terms.  Our result reveals the limitation of the venerable hydrogenic model of excitons, and highlights the importance of the Berry curvature in the effective mass approximation.
\end{abstract}


\maketitle

The effective mass approximation provides a simple yet extremely useful tool to understand a wide variety of electronic properties of semiconductors~\cite{fundamentals}.  Within this approximation, electrons behave almost like free particles in response to external fields, provided that one replaces the bare electron mass with an effective mass derived from the band dispersion.  Much of our intuition on electron transport is based on this semiclassical picture.  However, it has been shown that such a picture is actually incomplete, and one must include the Berry curvature of the Bloch states~\cite{xiao2010}.  Essentially, the Berry curvature modifies the electron dynamics through an anomalous term in the group velocity of the Bloch electrons~\cite{karplus1954,sundaram1999}, i.e.,
\begin{equation} \label{velocity}
\dot{\bm r} = \frac{1}{\hbar}\dpar{\eps_n(\bm k)}{\bm k} + \bm\nabla V(\bm r) \times \bm\Omega_n(\bm k) \;,
\end{equation}
where $\eps_n(\bm k)$ is the band energy, $V(\bm r)$ is the external potential, and $\bm\Omega_n(\bm k) = i\bracket{\bm\nabla_{\bm k}u_{n\bm k}|\times|\bm\nabla_{\bm k}u_{n\bm k}}$ is the Berry curvature defined in terms of the periodic part $u_{n\bm k}(\bm r)$ of the Bloch function.  The importance of the Berry curvature has been well established in a number of transport phenomena such as the anomalous Hall effect~\cite{fang2003,yao2004,nagaosa2010} and the spin Hall effect~\cite{murakami2003,sinova2004,culcer2004}.

In this Letter we consider another type of problems for which the effective mass approximation must be modified to include the Berry curvature, namely, the bound state problem of Bloch electrons.  To be specific, we will consider the energy spectrum of an exciton, even though our result should be equally applicable to other problems such as shallow impurity states.  Our motivation is two fold.  First, giant exciton binding energies (about a few hundred meV) have recently been observed in monolayers of transition metal dichalcogenides~\cite{mak2010,splendiani2010,mak2012,ross2013,ugeda2014,he2014,chernikov2014,ye2014,zhang2014,wang2015}, in which the low-energy carriers behave like massive Dirac fermions with nonzero Berry curvature~\cite{xiao2012}.  Thus, the detailed experimental study of excitons in the presence of the Berry curvature appears to be feasible.  Secondly, there have been a few calculations of the exciton energy spectrum in these materials~\cite{cheiwchanchamnangij2012,ramasubramaniam2012,qiu2013,ye2014,berghauser2014,wu2015,berkelbach2015}, but the role of the Berry curvature is not explicitly discussed.  We will show that, at the level of the effective mass approximation, the Berry curvature is essential to understand the exciton energy spectrum.

Our main results are summarized below.  We show that, quite generally, the Berry curvature modifies the effective Hamiltonian for excitons, and causes an energy splitting between exciton states with opposite angular momentum.  This splitting is determined by the Berry curvature flux through the $\bm k$-space area spanned by the relative motion of the electron-hole pair in the exciton wave function.  We confirm this result by a detailed study of the massive Dirac fermion model in two dimensions, and show that the energy splitting can be understood as an effective spin-orbit coupling effect.  In addition, we also find a shift of the energy levels due to other ``relativistic'' terms.  Finally, the effective Hamiltonian approach is compared with a $\bm k$-space Hatree-Fock calculation, where the gauge-dependence of the angular momentum number is discussed.  Our study provides a clear explanation of the previously calculated exciton energy splitting~\cite{ye2014,wu2015,berkelbach2015}.  It reveals the importance of the Berry phase in exciton physics, and calls for a thorough investigation of its effect on interacting phenomena.

An exciton is a bound state of a conduction band electron and a valence band hole attracted to each other via the Coulomb interaction.  Within the effective mass approximation, the motion of an exciton can be decomposed into a center-of-mass motion and a relative motion.  The latter is governed by the following effective Hamiltonian~\cite{dresselhaus1956,elliott1957}
\begin{equation} \label{hydrogen}
H = \frac{\hat p^2}{2\mu} + V(\hat{\bm r}) \;.
\end{equation}
where $\mu^{-1} = m_e^{-1} + m_h^{-1}$ is the reduced mass, $\bm r = \bm r_e - \bm r_h$ is the relative coordinate, $\bm p$ is the canonical momentum of $\bm r$, and $V(\bm r)$ is the screened Coulomb interaction.  For two-dimensional (2D) systems with a central potential $V(r)$, the eigenstates $(n, m)$ can be labeled by the radial quantum number $n$ and the angular momentum $m$.  For definitiveness, in the following we set $V(r) = -\kappa/r$, then the Hamiltonian describes a 2D hydrogen problem.  Our result, however, is independent of the detailed from of $V(r)$.  The exciton binding energy of the 2D hydrogen model is given by
\begin{equation} \label{Eb}
\begin{split}
&E_{n,m} = - \frac{\mathcal R}{(n + |m| + 1/2)^2} \;, \\
&n = 0, 1, 2, \dots, \quad m = 0, \pm 1, \pm 2, \dots
\end{split}
\end{equation}
where $\mathcal R = \mu\kappa^2/2\hbar^2$ is the Rydberg energy.  We can see that the states $(n, \pm m)$ with opposite angular momentum are degenerate, a general consequence of time-reversal symmetry.

\begin{figure}
\includegraphics[width=6cm]{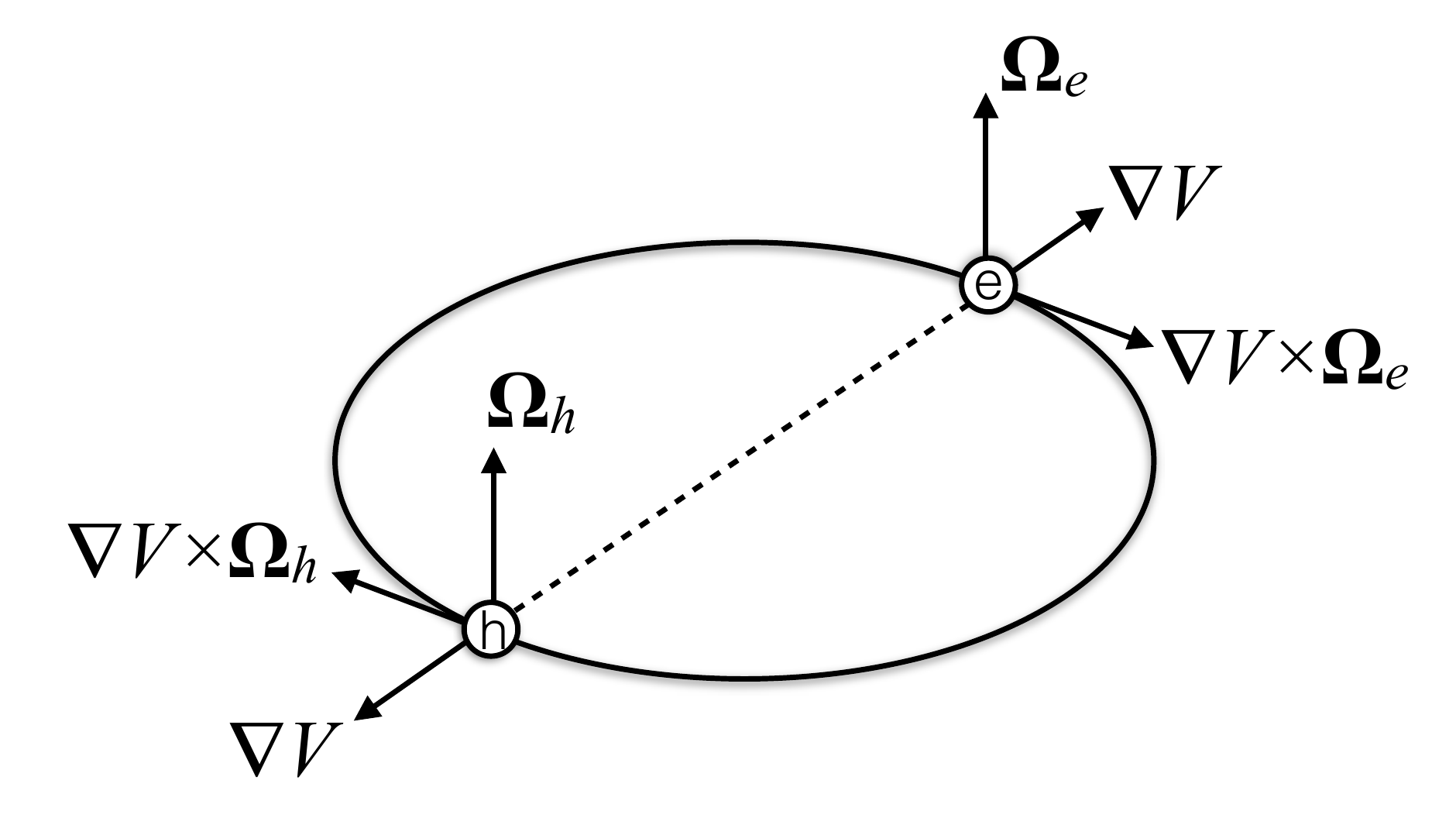}
\caption{\label{fig:pair}With finite Berry curvature ($\bm\Omega_e$ and $\bm\Omega_h$), the electron and the hole acquire an anomalous velocity $\bm\nabla V \times \bm\Omega$ in a central potential $V(r)$, resulting in a lift of the degeneracy between the left- and the right-rotating states.}
\end{figure}

The above picture is modified in the presence of the Berry curvature.  According to the semiclassical equation of motion~\eqref{velocity}, the electron and the hole will acquire an anomalous velocity perpendicular to the radial direction in a central potential (Fig.~\ref{fig:pair}).  Obviously, this anomalous term breaks time-reversal symmetry and should lead to an energy difference between the left- and the right-rotating states.

To obtain a quantitative theory of the energy spectrum, we need to quantize the semiclassical motion of the exciton.  This can be done using the canonical quantization procedure~\cite{chang2008,xiao2010}.  It has been shown that in the presence of the Berry curvature, the position operators become non-commutative and satisfy~\cite{fang2003,murakami2003,xiao2005},
\begin{equation}
[\hat r_\alpha, \hat r_\beta] = i\eps_{\alpha\beta\gamma}\Omega_\gamma \;.
\end{equation}
For the relative motion of the electron-hole pair, $\bm\Omega = \bm\Omega_e + \bm\Omega_h$ should be understood as the sum of the Berry curvatures of the electron and the hole~\footnote{The Berry curvature introduced here should not be confused with the exciton Berry curvature defined in Ref.~\cite{yao2008a}}.  In general, $\bm\Omega$ is a function of $\bm k$.  However, if the exciton wave function is sharply localized in the $\bm k$-space, then we can approximate $\bm\Omega$ with its value at the band edge.  To derive the effective Hamiltonian, we introduce the canonical coordinates~\cite{chang2008}
\begin{equation} \label{peierls}
\hat{\bm R} = \hat{\bm r} - \frac{1}{2\hbar}\bm\Omega\times\hat{\bm p} \;.
\end{equation}
Equation~\eqref{peierls} can be viewed as the $\bm k$-space counterpart of the Peierls substitution.  One can verify that, to first order, $[\hat R_\alpha, \hat R_\beta] = 0$.  Inserting Eq.~\eqref{peierls} back into Eq.~\eqref{hydrogen} and expanding to the first order in $\bm\Omega$, we obtain the effective Hamiltonian casted in the canonical variables,
\begin{equation} \label{heff}
H = \frac{\hat p^2}{2\mu} + V(\hat R) + \frac{1}{2\hbar}\bm\Omega \cdot (\bm\nabla V \times \hat{\bm p}) \;.
\end{equation}
Clearly, the extra term proportional to the Berry curvature will split the exciton states with opposite angular momentum.  

Equation~\eqref{heff} is the main result of our paper.  To gain some physical intuition, we apply our theory to the 2D hydrogen model.  The energy splitting between the exciton states $(0, \pm 1)$ is found to be
\begin{equation} \label{split}
\Delta E = \frac{64}{81}\frac{\Omega}{a_0^2}\mathcal R \;,
\end{equation}
where $a_0 = \hbar^2/\kappa\mu$ is the Bohr radius of the exciton envelop function.  Since $a_0^{-2}$ is roughly the $\bm k$-space spread of the envelop function, one can interpret the energy splitting~\eqref{split} as proportional to the $\bm k$-space Berry phase flux penetrating the area occupied by the exciton.  In 2D systems the screened Coulomb interaction $V(r)$ has a rather complicated $r$-dependence, which leads to a non-hydrogenic Rydberg series of exciton states~\cite{chernikov2014,ye2014}.  Nonetheless, the energy splitting, and its interpretation in terms of the Berry phase flux, are independent of the detailed form of $V(r)$.

To further demonstrate the above physics, in the following we turn to a concrete model, i.e., the gapped 2D Dirac equation,
\begin{equation} \label{dirac}
H_0 = \alpha \bm k \cdot \bm\sigma + \Delta\sigma_z \;,
\end{equation}
where $\bm k = (k_x, k_y)$ is the 2D wave vector, and $2\Delta$ is the band gap.  This Hamiltonian describes the low-energy carriers in a number of materials, including topological surface states~\cite{garate2011,efimkin2013} where $\bm\sigma$ refers to the electron spin, and gapped graphene~\cite{xiao2007} and semiconducting dichalcogenides~\cite{xiao2012} where $\bm\sigma$ refers to the orbital index.  The energy dispersion is given by $\eps_{c,v} = \pm \eps_{\bm k} = \pm\sqrt{\Delta^2 + \alpha^2k^2}$ with the corresponding eigenstates
\begin{equation} \label{basis}
\ket{c\bm k} = \binom{\cos\frac{\theta_{\bm k}}{2}}{e^{i\phi_{\bm k}}\sin\frac{\theta_{\bm k}}{2}} \;, \quad
\ket{v\bm k} = \binom{e^{-i\phi_{\bm k}}\sin\frac{\theta_{\bm k}}{2}}{-\cos\frac{\theta_{\bm k}}{2}} \;,
\end{equation}
where the subscript $c$ and $v$ label the conduction and valance bands, respectively, and the angular variables $\theta_{\bm k}$ and $\phi_{\bm k}$ are defined as $\theta_{\bm k} = \cos^{-1}(\Delta/\eps_k)$ and $\phi_{\bm k} = \tan^{-1}(k_y/k_x)$.  The Berry curvature is given by~\cite{xiao2007,xiao2012}
\begin{equation} \label{curvature}
\Omega_e = \Omega_h = -\frac{\alpha^2\Delta}{2(\Delta^2 + \alpha^2k^2)^{3/2}} \;.
\end{equation}
Note that the Berry curvature of holes should be opposite to that of the valence band electrons.  The fact that $\Omega_e = \Omega_h$ is a specific feature of the two-band model.  In general $\Omega_e$ and $\Omega_h$ are different.  In the following we are going to approximate the Berry curvature with its $\bm k = 0$ value, and define the joint Berry curvature $\Omega = \Omega_e(0) + \Omega_h(0) = -\alpha^2/\Delta^2$ (assuming $\Delta > 0$).

We now estimate the energy splitting of excitons in transition metal dichalcogenides.  The band structure parameters $\alpha$ and $\Delta$ have been calculated in Ref.~\cite{xiao2012}.  For all four compounds $MX_2$ ($M = $ Mo, W, and $X = $ S, Se), $\Omega$ is about 15 \AA$^2$.  The Bohr radius of the $s$-state exciton is $a_0 \sim 20$ \AA~\cite{ye2014}.  Hence the energy splitting $\Delta E$ between the two $p$-states is roughly 4\% of the exciton binding energy, or a few tens of meV.  This is consistent with previous calculations based on solving the Bethe-Salpeter equation~\cite{ye2014,wu2015}.  In dichalcogenides, the band structure consists of two valleys located at the two inequivalent corners of the hexagonal Brillouin zone with opposite Berry curvature.  Therefore, the energy splitting in the two valleys are opposite, restoring the overall time-reversal symmetry of the system.
 
Since the 2D Dirac equation~\eqref{dirac} also describes relativistic spin-$\frac{1}{2}$ particles, it is interesting to explore the connection between our result and relativistic quantum mechanics.  In the latter case, the effective Hamiltonian, also known as the Schr\"odinger-Pauli equation, is usually obtained using the Foldy-Wouthuysen transformation~\cite{foldy1950}.  The application to the 2D Dirac equation parallels exactly to its 3D counterpart.  After adopting the center-of-mass and the relative coordinates~\footnote{For details, see the derivation of Eq.~(2) in Ref.~\cite{garate2011}},
We find the effective Hamiltonian for the positive energy branch is
\begin{equation}
H_\text{eff} = \frac{\hat p^2}{2\mu} + V(R) + \frac{1}{2\hbar}\bm\Omega \cdot (\bm\nabla V\times \hat{\bm p}) + \frac{1}{4}\Omega\nabla^2V \;,
\end{equation}
where $\mu = \hbar^2\Delta/2\alpha^2$ is the reduced mass, and $\bm\Omega = \Omega\hat{\bm e}_z$.  It is now clear that the Berry-curvature caused splitting stems from an effective spin-orbit coupling term (the third term).  We also note that there is an extra term proportional to $\nabla^2V$, known as the Darwin term~\cite{relativity}.  This term does not appear in the semiclassical quantization scheme because the semiclassical formalism is only accurate to the first order of $\bm\nabla V$~\cite{chang2008}.  The Darwin term will lead to an energy shift depending on the radial quantum number $n$~\footnote{We note that there is a difference between 2D and 3D regarding the Darwin term.  In 3D, for the Coulomb interaction $\nabla^2 V \propto \delta (r)$, therefore the Darwin term only affects the energy levels of $s$-states.  In 2D this is no longer true; the Darwin term can affect states with nonzero angular momentum as well.}.  However, given the central symmetry of $V(r)$, the energy shift of states $(n, \pm m)$ are the same.  Therefore the energy splitting between these two states are entirely due to the Berry curvature effect discussed above.  One can also carry out the Foldy-Wouthuysen transformation to higher orders, which only leads to quantitative changes.

It is useful to compare the effective Hamiltonian approach to the $\bm k$-space formalism based on the Bethe-Salpeter equation.  An exciton at rest can be written as $\ket{\Phi_\text{ex}} = \sum_{\bm k} f(\bm k) a^\dag_{c\bm k} a_{v\bm k}\ket{\Phi_0}$, where $\ket{\Phi_0}$ is the ground state in which all valence bands are filled and all conduction bands are empty, and $a_{c\bm k}$ and $a_{v\bm k}$ are the annihilation operators for the conduction and valence band electrons, respectively.  Following the standard procedure~\cite{garate2011,wu2015}, the exciton Hamiltonian for the envelop function $f(\bm k)$ is given by
\begin{equation}
(2\eps_{\bm k} + \Sigma_{\bm k})f(\bm k) - \sum_{\bm k'} U(\bm k, \bm k') f(\bm k') = E f(\bm k) \;,
\end{equation}
where $\Sigma_{\bm k}$ is the self energy, which we shall absorb into the definition of the optical gap and will not be written explicitly hereafter, and $U(\bm k, \bm k')$ is the Coulomb interaction between the electron and the hole,
\begin{equation}
U(\bm k, \bm k') = V(\bm k-\bm k')\bracket{c\bm k|c\bm k'}\bracket{v\bm k'|v\bm k} \;,
\end{equation}
with $V(\bm k-\bm k')$ the Fourier transform of the Coulomb interaction $V(r)$.  We note that $\bracket{c\bm k|c\bm k'}$ can be written as $|\bracket{c\bm k|c\bm k'}|e^{i\Delta\phi}$, where $\Delta\phi = \IM\log\bracket{c\bm k|c\bm k'}$ is the discretized Berry phase~\cite{resta2000}.  It is through this term that the Berry curvature enters the picture.  

For the isotropic system considered here, we can decompose the envelop function into different angular momentum channels, i.e., $f_m(k) = \int_0^{2\pi} \frac{d\phi}{2\pi}\, f(\bm k) e^{im\phi_{\bm k}}$.  The corresponding equation is
\begin{equation} \label{exciton}
2\eps_k f_m(k) - \int_0^\infty \frac{k'dk'}{2\pi}U_m(k, k')f_m(k') = E f_m(k) \;,
\end{equation}
where the Coulomb interaction for the $m$-th channel is given by $U_m(k, k') = \int_0^{2\pi} \frac{d\phi}{2\pi} e^{im\phi} U(\bm k - \bm k')$, $\phi = \phi_k - \phi_{k'}$ is the relative angle between $\bm k$ and $\bm k'$.

Before moving on, we comment on the physical meaning of $m$.  Note that the matrix element $U(\bm k - \bm k')$ is actually gauge dependent.  Different choices of the basic function $\ket{c\bm k}$ and $\ket{v\bm k}$ will not change the energy spectrum, but will lead to an integer shift of $m$.  Specifically, if we perform a gauge transform $\ket{\lambda\bm k} \to e^{i\phi_\lambda(\bm k)} \ket{\lambda\bm k}$, then the Coulomb interaction becomes
\begin{equation}
U(\bm k, \bm k') \to e^{i[\phi_c(\bm k') - \phi_c(\bm k)]} e^{-i[\phi_v(\bm k') - \phi_v(\bm k)]} U(\bm k, \bm k') \;.
\end{equation}
Consequently, the Coulomb interaction in the $m$-th channel will be shifted to the $(m+n)$-th channel in the transformed basis, where $n$ is the winding number of $\phi_\lambda(\bm k)$.  Therefore one should be careful when labeling the exciton states using $m$.  There seems to be some confusion over this fact in the literature~\cite{stroucken2014}.  We have chosen the basis function~\eqref{basis} such that in the limit of vanishing Berry curvature, the labeling of $m$ returns to that of the 2D hydrogen model.

We can now expand the Coulomb interaction to the leading order of  $\alpha/\Delta$ and obtain
\begin{equation} \label{Vmapp}
U_m(\bm k,\bm k')\approx V_m (\bm k - \bm k') + \frac{\alpha^2kk'}{2\Delta^2}V_{m+1}(\bm k-\bm k') \;.
\end{equation}
Clearly, $U_m$ and $U_{-m}$ differ by a term proportional to $\alpha^2/\Delta^2$, which is nothing but the joint Berry curvature $\Omega$.  We also note that the energy shift of $\pm m$ states is asymmetric.  This is caused by an overall shift of both $\pm m$ states as mentioned earlier.
\begin{figure}
\includegraphics[width=\columnwidth]{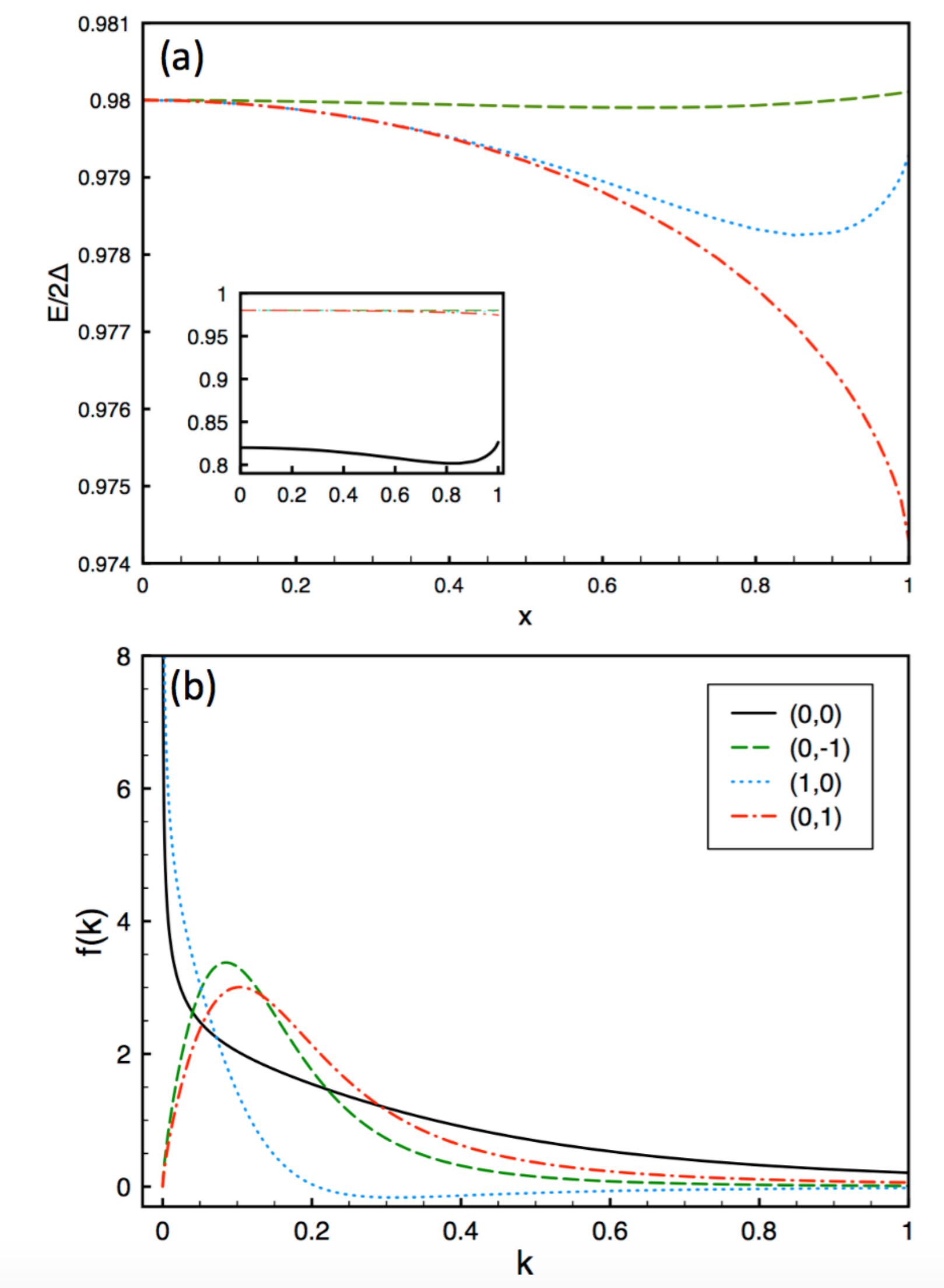}
\caption{\label{fig:energy} (Color online) (a) Exciton energy spectrum as a function of the inter-band coupling strength $x$.  The insert shows all four states $(0, 0)$, $(1, 0)$, and $(0, \pm 1)$.   (b) Exciton wave function at $x = 1$.}
\end{figure}

Finally, to demonstrate the essential role of the Berry curvature, we consider a modified Hamiltonian
\begin{equation}
\tilde H = x\alpha \bm k \cdot \bm\sigma + (\Delta + \beta_x k^{2}) \sigma_{z} \;,
\end{equation}
where $\beta_x = (1-x^2)\alpha^2/2\Delta$.  Here the parameter $x$ can be regarded as a measure of the inter-band coupling strength.  One can verify that as $x$ changes from 0 to 1, the effective mass stays the same, $m^* = \hbar^2\Delta^2/\alpha^2$, whereas the Berry curvature gradually increases from 0 to its value given in Eq.~\eqref{curvature}.  

We numerical solve Eq.~\eqref{exciton} with $V(r) = -\kappa/r$ using the modified Gauss-Legendre quadrature method with a constant scaling~\cite{garate2011,chao1991}.  Figure~\ref{fig:energy}(a) shows the calculated exciton energy spectrum as a function of $x$.  At $x = 0$, the Berry curvature is zero, and the three states $(0, \pm 1)$ and $(1, 0)$ are degenerate as indicated by Eq.~\eqref{Eb}.  As $x$ increases, the energy difference between $(0, \pm 1)$ starts increasing and reaches its maximum at $x = 1$, when the Berry curvature is also maximal.  The asymmetric splitting is obvious.  Figure~\ref{fig:energy}(b) shows the $k$-space exciton wave functions, which clearly display the characteristic shape for $s$- and $p$-states, respectively.  This confirms our choice of the basis function.  We can see that there is a slight difference between $(0, \pm 1)$ states as a result of the Berry curvature.  The nonmonotonic behavior of the $m = 0$ states is due to the competition between the Darwin term and a higher order term proportional to $\hat p^4$ in the Foldy-Wouthuysen transformation~\cite{foldy1950}.

In summary, we have demonstrated that in the presence of the Berry curvature, the effective mass approximation must be modified through a $\bm k$-space Peierls substitution.  This results in a Berry-curvature induced energy splitting of exciton states with opposite angular momentum, which can be understood as an effective spin-obit coupling effect.  The method outlined in this paper is quite general, and can be easily transferred to other bound-state problems such as shallow impurity states.

We acknowledge useful discussions with Ming-Che Chang,  Shun Lien Chuang, Ion Garate,  Tony Heinz, Qian Niu, Junren Shi, and Xiaodong Xu.  This work is supported by DOE Basic Energy Sciences No.~DE-SC0012509 (D.X.\ and W.S.) and by AFOSR No.~FA9550-14-1-0277 (J.Z.)

\textit{Note added}.---Upon the completion of this work, we become aware of Ref.~\cite{srivastava2015}, in which the relation between exciton energy splitting and the Berry phase is discussed using the $k$-space formulation.

\end{document}